\documentclass{jpp}

\usepackage{graphicx}
\usepackage{natbib}

\ifCUPmtlplainloaded \else
  \checkfont{eurm10}
  \iffontfound
    \IfFileExists{upmath.sty}
      {\typeout{^^JFound AMS Euler Roman fonts on the system,
                   using the 'upmath' package.^^J}%
       \usepackage{upmath}}
      {\typeout{^^JFound AMS Euler Roman fonts on the system, but you
                   dont seem to have the}%
       \typeout{'upmath' package installed. JPP.cls can take advantage
                 of these fonts, if you use 'upmath' package.^^J}%
      }
  \else
  \fi
\fi

\ifCUPmtlplainloaded \else
  \checkfont{msam10}
  \iffontfound
    \IfFileExists{amssymb.sty}
      {\typeout{^^JFound AMS Symbol fonts on the system, using the
                'amssymb' package.^^J}%
       \usepackage{amssymb}%

      }{}
  \fi
\fi

\ifCUPmtlplainloaded \else
  \IfFileExists{amsbsy.sty}
    {\typeout{^^JFound the 'amsbsy' package on the system, using it.^^J}%
     \usepackage{amsbsy}}
    {}
\fi

\newsavebox{\astrutbox}
\sbox{\astrutbox}{\rule[-5pt]{0pt}{20pt}}

\title[Laboratory plasma physics experiments using merging
supersonic plasma jets]{Laboratory plasma physics experiments using merging supersonic plasma jets}

\author[S. C. Hsu et al.]
{S. C. Hsu,$^1$\thanks{Email: scotthsu@lanl.gov}  A. L. Moser,$^1$\thanks{Now at General Atomics, San Diego, CA.}
E. C. Merritt,$^{1,2}$
C. S. Adams,$^{1,2}$ J. P. Dunn,$^1$ S. Brockington,$^3$ A. Case,$^3$ M. Gilmore,$^2$ A. G. Lynn,$^2$ 
S. J. Messer,$^3$\thanks{Now at Foreground Security, Herndon, VA.} and F. D. Witherspoon$^3$}
\affiliation{$^1$Los Alamos National Laboratory, Los Alamos, NM 87545, USA\\[\affilskip]
$^2$University of New Mexico, Albuquerque, NM 87131, USA\\[\affilskip]
$^3$HyperV Technologies Corp., Chantilly, VA  20151, USA}

\pubyear{2014}
\volume{}
\pagerange{}
\date{?; revised ?; accepted ?. - To be entered by editorial office}
\begin{document}

\maketitle

\begin{abstract}
We describe a laboratory plasma physics experiment at Los Alamos National Laboratory that uses
two merging supersonic plasma jets formed and launched by pulsed-power-driven railguns.
The jets can be formed using any atomic species or mixture available in a compressed-gas bottle and
have the following
nominal initial parameters at the railgun nozzle exit:  $n_e\approx n_i \sim 10^{16}$~cm$^{-3}$,
$T_e \approx T_i \approx 1.4$~eV, $V_{\rm jet}\approx 30$--100~km/s, mean charge $\bar{Z}\approx 1$, 
sonic Mach number $M_s\equiv V_{\rm jet}/C_s>10$, jet diameter $=5$~cm,
and jet length $\approx 20$~cm.  Experiments to date have
focused on the study of merging-jet dynamics and the shocks that form as a result of the
interaction, in both collisional and 
collisionless regimes with respect to the inter-jet classical ion mean free path, and with and without an
applied magnetic field.
However, many other studies are also possible, as discussed in this paper.  
\end{abstract}

\begin{PACS}
%Authors should not enter PACS codes directly on the manuscript, as these must be chosen during the online submission process and will then be added during the typesetting process (see http://www.aip.org/pacs/ for the full list of PACS codes)
\end{PACS}

\section{Introduction}
\label{sec:introduction}

The Plasma Liner Experiment (PLX)
is a laboratory plasma physics facility at Los Alamos National 
Laboratory.  The primary purpose and goal of PLX is to form and study spherically imploding
plasma liners via multiple merging plasma jets, motivated by
high-energy-density (HED) physics and magneto-inertial fusion (MIF) applications
\citep{thio99,thio01,hsu12a,hsu12b, santarius12,knapp14}.
For fundamental
HED-physics studies \citep{drake06}, PLX is envisioned to provide an economic means for forming 
inertially confined cm-, $\mu$s-, and Mbar-scale plasmas \citep{awe11,davis12,cassibry13} upon the stagnation of up 
to 60 merged pulsed-power-driven, supersonic
plasma jets.  Such an experiment would allow for good diagnostic measurements on larger-spatial and 
longer-temporal scales and a much higher shot rate than those of typical laser- or Z-pinch-driven HED experiments.
For the MIF application \citep{lindemuth83,kirkpatrick95,lindemuth09a},
the use of a spherically converging plasma liner for compressing a magnetized target
plasma would avoid the repetitive hardware destruction inherent in solid-liner approaches \citep[e.g.,][]{intrator04,slutz10,degnan13}, and it would allow for a higher
repetition rate ($\sim 1$~Hz), which is desirable for an economic, pulsed fusion-energy system.

The PLX facility was built in 2010--2011, and the first plasma jet was fired in September, 2011.  Initial research
characterized in detail the plasma properties and evolution of a single jet \citep{hsu12b}.  A second railgun was
installed in mid-2012, and ensuing experiments focused on the oblique merging of two plasma jets, resulting
in detailed observations shown to be consistent with collisional plasma shocks \citep{merritt13,merritt14}.  Due to 
loss of funding, we were not able to build the facility up to 30 railguns as originally planned.  Instead,
ongoing experiments are focused on the head-on merging of two plasma jets
to form and study unmagnetized and magnetized plasma shocks.  An in-chamber
pulsed Helmholtz coil was installed in late 2013 to enable the study of magnetized shocks.
The facility now enables a unique
experimental research program on the detailed study of plasma shocks, from collisional-to-collisionless
and unmagnetized-to-magnetized regimes.  Preliminary designs for a 36- or 60-jet experiment (with
$\sim 1.5$~MJ of capacitive stored energy)
to form spherically imploding plasma liners are also in hand.

Plasma jet and shock studies on PLX complement many other related, contemporary experimental efforts.  Examples
include studies of plasma jet dynamics, interactions, and/or shock formation in astrophysically or fusion-relevant
contexts using a variety of plasma formation methods, ranging from coaxial plasma guns
\citep[e.g.,][]{bellan05,hsu05,moser12}
to wire-array-driven pinches \citep[e.g.,][]{haas11,gourdain13,swadling14} to laser-driven experiments \citep[e.g.,][]{ross13,li13,fox13,fiksel14}.  In general, PLX plasmas are colder and more collisional than the other experiments,
but have the benefits of larger spatial size for ease of diagnostic measurements, more choices for the working
plasma species, and the option of both unmagnetized
and magnetized experiments.  A brief overview of many plasma-jet experiments and formation methods is
given in \citet{hsu09}.

The rest of this paper is organized as follows.  Section~\ref{sec:description} describes the PLX experimental device
and parameters.  Section~\ref{sec:physics} discusses the physics questions that can be addressed on PLX\@.  
Section~\ref{sec:results} summarizes the major results obtained thus far, and Sec.~\ref{sec:opportunities}
describes future opportunities.

\section{Description of experimental device and parameters}
\label{sec:description}

\subsection{Device description}
\label{sec:device}
\begin{figure}
\centerline{\includegraphics[width=3truein]{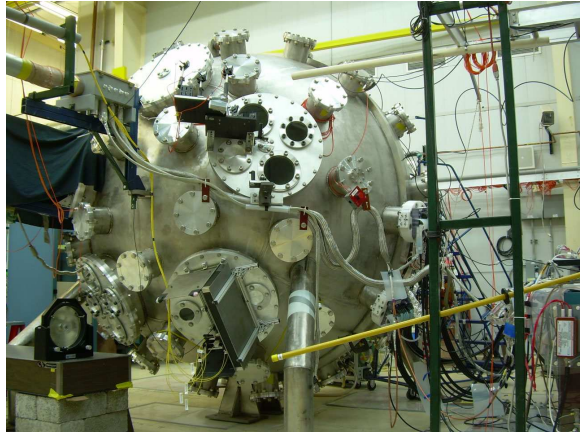}}
\caption{Photograph of the PLX vacuum chamber (2.74-m diameter).}
\label{fig:plx}
\end{figure}

PLX consists of a 9-ft. (2.74-m) diameter spherical vacuum chamber (Fig.~\ref{fig:plx}), two pulsed-power-driven 
plasma railguns (Sec.~\ref{sec:railguns}), a diagnostic suite (Sec.~\ref{sec:diagnostics}),
and an in-chamber Helmholtz coil (Sec.~\ref{sec:coils}) to generate a pulsed magnetic field up to 0.22~T\@.  
The facility is housed in a 3000-ft.$^2$ (279-m$^2$) high-bay space
with a 10-ton (9072-kg) bridge crane, roll-up door with 18.5-ft.\ (5.6-m) tall and 16-ft.\ (4.9-m) wide clearance,
208- and 480-VAC power, and building chilled water and compressed air.

\subsubsection{Plasma railguns and pulsed-power systems}
\label{sec:railguns}

\begin{figure}
\centerline{\includegraphics[width=5.3truein]{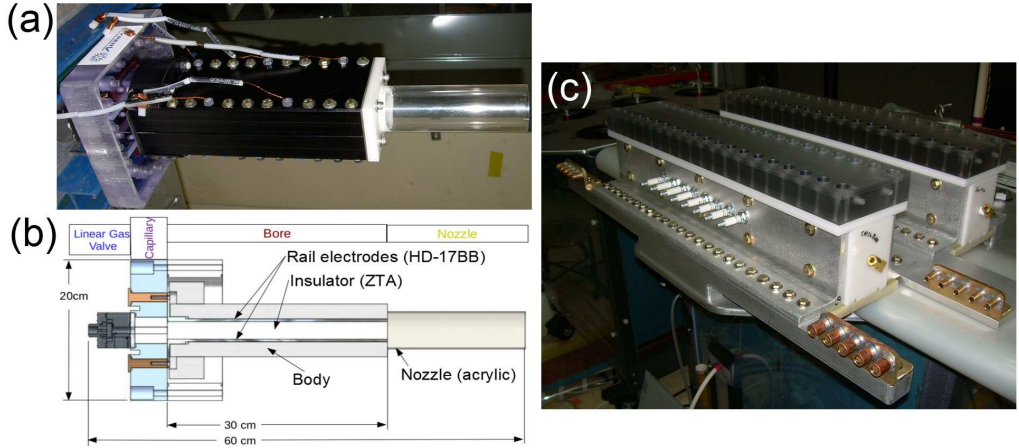}}
\caption{(a)~Photograph of railgun.  (b)~Schematic of railgun (courtesy
of HyperV Technologies).  (c)~Photograph of two linear,
spark-gap switches (1 main switch, 1 crowbar) installed on top of a railgun capacitor bank.}
\label{fig:gun-switch}
\end{figure}

The plasma railguns [Fig.~\ref{fig:gun-switch}(a) and (b)] and their high-current, spark-gap switches [Fig.~\ref{fig:gun-switch}(c)]
were designed and built by collaborator HyperV Technologies Corp.  The railgun
has two parallel-plate electrodes made of tungsten alloy (HD-17BB),
separated by insulators made of zirconium-toughened alumina (ZTA)\@.
The railgun bore is 1~in.\ (2.54~cm) $\times$ 1~in.\ (2.54~cm),
and the body is a clamshell design consisting of two Noryl halves bolted
together [Figs.~\ref{fig:gun-switch}(a) and (b)].  
A custom-built gas-puff valve injects neutral gas into a pre-ionization (PI) chamber at the
rear of the railgun.  A cylindrical acrylic nozzle with 5-cm inner diameter is mounted at the exit of the railgun bore.
The rails, PI, and gas valves (GV) are driven by four separate capacitor banks, which are
summarized in Table~\ref{tab:cap-banks}.  The GVs and PIs of both guns are connected in parallel and driven by a
single GV and PI bank, respectively.

%\subsection{Tables}
%Tables, however small, must be numbered sequentially in the order in which they are mentioned in the text. The word 
%\textit {table} is only capitalized at the start of a sentence. See table \ref{tab:kd} for an example.

\begin{table}
 \begin{center}
\def~{\hphantom{0}}
 \begin{tabular}{lcccc}
& capacitor & rated bank & total bank & max. \\
& model & voltage (kV) & capacitance ($\mu$F) & energy (kJ) \\[3pt]
Railguns (one bank per gun) & Maxwell 32184 & 40 & 12 or 36 & 9.6 or 28.8\\
%Railgun 2 & Maxwell 32814 & 40 & 12 or 36 & 9.6 or 28.8\\
Pre-ionization (for both guns) & Maxwell 32814 & 40 & 12 & 9.6 \\
Gas valve (for both guns) & Maxwell 32567 & 20 & 24 & 1.7 \\
Helmholtz coil & GA 32934 & 5 & 4000 & 50\\
  \end{tabular}
  \caption{Capacitor bank parameters.}
  \label{tab:cap-banks}
  \end{center}
\end{table}

\subsubsection{Diagnostics}
\label{sec:diagnostics}

Operational diagnostics include a multi-chord visible interferometer, a visible-to-near-infrared survey spectrometer,
an array of three photodiodes, a charge-coupled-device (CCD) camera, a schlieren imaging system,
an electrostatic triple probe, and magnetic probe arrays.  We briefly summarize the diagnostic capabilities here;
more technical details are given elsewhere \citep{hsu12b,lynn10}.

The 561-nm multi-chord interferometer \citep{merritt12a,merritt12b} is fiber-coupled between the launch laser optics
and the vacuum chamber, allowing for relatively easy (a few days work) modification of chord positioning
for different experiments and different plasma gun geometries.   
Several different chord arrangements have been used for single-jet propagation \citep{hsu12b}, two-jet 
oblique-merging experiments \citep{merritt13,merritt14}, and two-jet head-on merging experiments \citep{moser14}.
The time-resolution of the continuous
interferometer phase signals is 50~ns, limited by the
40-MHz digitization rate.  Spatial resolution is determined by the chord spacing (typically a few cm) 
and the laser spot size ($\approx 3$-mm diameter at the position of the plasma).

The survey spectrometer system consists of a 0.275-m spectrometer and a gated (0.45-$\mu$s duration),
1024-pixel, multi-channel-plate array covering the wavelength range 300--900~nm.  The best 
spectral resolution, corresponding to a 600~line/mm grating, is 0.152~nm/pixel.  
Plasma light is collected at a vacuum chamber window into
a 5-mm diameter
collimating lens and brought to the spectrometer with a optical fiber bundle.  The diameter
of the spectrometer viewing chord is $\approx 7$~cm at the position of the plasma.
The lens and fiber can be moved 
easily to different viewing positions.

A three-channel photodiode (PD) array collects chord-integrated and time-resolved
broadband visible (300--850~nm) plasma emission for determining jet
propagation speed, based on the viewing-chord positions and time-of-flight estimates of features
in the PD data traces.  The peak PD responsivity is 0.65~A/W at 970~nm, and the frequency response
decreases with increasing gain (2.1~MHz and 0.1~MHz at 20 and 50~dB, respectively).  The light is
collected through an adjustable aperture that is typically set at 1~cm, which constitutes the nominal spatial
resolution.

An intensified CCD camera (DiCam Pro ICCD), with spectral sensitivity from the UV to near-IR, is used
to capture time-gated, visible images of the experiment.  The camera records $1280\times1024$-pixel
images with 12-bit dynamic range.  The typical gate time used is 20~ns.  The camera is triggered remotely
via an optical fiber, housed inside a metal shielding box, and mounted next to a large rectangular
borosilicate window on the vacuum chamber (although it can be moved to other ports as well).

A schlieren system \citep{settles01}
for obtaining 2D images (4~cm $\times$ 16~cm) sensitive to plasma-density gradients has been constructed.
The PLX schlieren system \citep{adams12}
uses a pulsed 1.064-$\mu$m Nd:YAG laser (3-ns pulse, 30--80~mJ) expanded to a 20-cm diameter beam
that is then collimated and passed through the plasma between 25-cm diameter, $f$/8 mirrors.  A tunable knife
edge blocks light refracted by density gradients in the plasma.  The image is captured by an IR camera
(Apogee Alta U1109) which contains a $2048 \times 512$ CCD detector with 8\% quantum efficiency at
1.064 $\mu$m.  The system is designed to give images with $>5$\% contrast for electron-density gradients
$\gtrsim 10^{15}$~cm$^{-4}$, i.e., for 1-cm-scale gradients in the presence of $10^{14}$-cm$^{-3}$ density 
plasmas.  The intensity of the recorded image is proportional to $\nabla N$, where $N$ is the refractive
index of the plasma.  The PLX schlieren diagnostic has yet to be tested on plasmas with sufficient density gradients
to yield images with detectable contrast.

Magnetic probe arrays measure all three components of the
magnetic field along the railgun nozzle and
within the plasma-jet-interaction region.  The nozzle probe array consists of hand-wound coils (using 0.5-mm
diameter Kapton-coated magnet wire) around a rectangular Delrin rod ($0.37\times0.48$~cm$^2$).  These
probes have effective turns $\times$ area ($NA$) of $\approx 10$~cm$^2$ in the frequency
range 10--5000~kHz.  The internal probe array is mounted
inside an insertable probe shaft and contains two sets of 3-axis coils at two positions separated by 27.75~cm, for
the simultaneous measurement of up to six $B$-dot signals simultaneously.  These probe arrays use
commercial inductor chips (Coilcraft 1008CS-272XJLB, with dimensions $2.92\times 2.79\times 2.03$~mm$^3$);
the probe construction is similar to that described by \citet{romero04}.  The internal probe array coils have
$NA\approx 0.7$~cm$^2$ between 10--1000~kHz.  The $B$-dot signals are integrated by a passive integration
circuit with $RC=100$~$\mu$s.

An electrostatic triple probe \citep[e.g.,][]{ji91}
measures the instantaneous floating potential $V_f$, electron
temperature $T_e$, and ion saturation current $I_{\rm sat}$ as a function of time at one position in space.  The
probe can be moved to different radial positions with respect to chamber center.  Due to the lack of a
reliable, nearby reference ground potential (due to large voltage excursions
on the vacuum chamber during a shot), the triple probe is not yet producing reliable data.  To improve
the probe operation, we have built
4 battery-powered, analog optical channels (up to 1-MHz frequency response) to bring probe signals back to the
data acquisition system.

Finally, there are Rogowski coils, high-voltage probes, voltage divider networks, and pulsed current
monitors to record the pulsed
electrical currents and voltages of all the capacitor banks.

\subsubsection{Control and data acquisition systems}
\label{sec:control-daq}

PLX has an FPGA-based control/trigger system and CAMAC-based digital-data-acquisition system, all integrated
through a LabVIEW software interface and operated from a desktop computer.  Both analog and optical channels are
available for diagnostic triggering and operating relays for the various hardware sub-systems.  The data-acquisition system consists of 64 channels with 12-bit resolution, up to 40-MHz sampling rate,
and 32~kilosamples of memory per channel (4 Joerger TR modules).  Shot information and diagnostic data are
stored in an MDSplus tree (www.mdsplus.org).

\subsubsection{In-chamber Helmholtz coils}
\label{sec:coils}

We have installed inside the vacuum chamber
a pair of magnetic coils arranged in a Helmholtz (HH) configuration (coil radius
and separation $\approx 30$~cm)
inside the vacuum chamber.  
The coils and mounting structure
were designed and fabricated by Woodruff Scientific, LLC, and are presently installed such that the HH 
field is perpendicular to the direction of jet propagation, but
the HH field can be adjusted up to $\pm 30^\circ$ about the perpendicular and parallel directions with respect to the
jet propagation direction.  The coils are driven by a capacitor bank (see Table~\ref{tab:cap-banks}) with a current-rise
time of $\approx 1.3$~ms.  The maximum HH-bank
operating voltage of 4~kV gives a coil current $\approx 8.2$~kA
and a peak magnetic field of 0.22~T\@.  The minimum operating voltage is around 200~V (corresponding
to about 0.44 kA and 0.012~T peak magnetic field), below which the
HH-bank ignitron switch does not fire reliably.  The magnetic-field profiles have been characterized experimentally
using three concentric flux loops of different radii.

\begin{figure}
\centerline{\includegraphics[height=3truein]{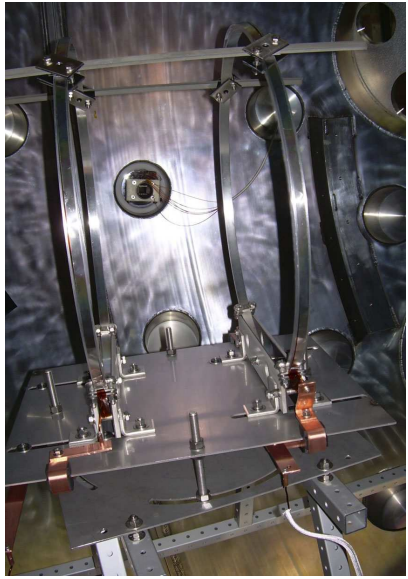}}
\caption{Photograph of the in-chamber Helmholtz coil, looking down the barrel of one railgun.}
\label{fig:coil}
\end{figure}

\subsection{Parameters}
\label{sec:parameters}

Table~\ref{tab:parameters}
summarizes the experimentally measured or inferred physical parameters of our plasma jets, both
at the gun-nozzle exit and after propagating $\approx 41$~cm.  
For head-on jet-merging experiments, the jets propagate $\approx 1.1$~m before colliding with each other
near chamber center, where the jet $n_{tot}\sim 10^{14}$~cm$^{-3}$ and $T_e\approx 1$~eV just before the collision.
For the achievable
range of $n_{tot}$, $T_e$, and $V$ just before jets collide, the inter-jet counter-streaming ion mean free path
$\lambda_{\rm i,ctr}$ (mean free path of an ion in one jet interacting with all species in the opposing jet)
can range from collisional, i.e., $\lambda_{\rm i,ctr} < \Delta$ \citep{merritt13,merritt14},
where $\Delta$ is the gradient-scale-length
of the shock that forms as a result of the colliding jets, to collisionless, i.e., $\lambda_{\rm i,ctr} \gg \Delta$ 
(the goal of ongoing work).
Note that both ion and electron collisionality are very high within each jet, and that the inter-jet
counter-streaming electron mean 
free path is very short, i.e., $\lambda_{\rm e,ctr}\ll \Delta$.

\section{Physics questions that can be addressed}
\label{sec:physics}

\subsection{Plasma shock structure and dynamics}

Highly supersonic plasmas, i.e., in which the sonic Mach number $M\equiv V/C_s \gg 1$ (with $V$ the plasma flow
speed and $C_s$ the ion sound speed), are needed and provide a natural opportunity to study plasma shock
physics \citep[e.g.,][]{jaffrin64,casanova91,sagdeev91}.  
Detailed laboratory studies of the latter have been scarce for two main, coupled reasons:  
(i)~the most readily available ways to form supersonic plasmas tended to be in very high-density regimes, e.g., using
laser, $z$-pinch, or $\theta$-pinch sources that (ii)~presented significant diagnostic challenges.  The latter
arose due to a combination of small spatial scales, high plasma densities, and/or lack of diagnostic access.
PLX, with its macroscopic 
plasma jets at moderate density and completely open diagnostic access, presents a unique experimental
platform for studying detailed plasma shock structure and dynamics on multiple-cm spatial and several-$\mu$s
time scales.  The experiment also offers the ability to span collisional to collisionless
regimes for the counter-streaming ions and unmagnetized to magnetized jet interactions.

\subsection{Equation-of-state in regimes intermediate between magnetic and inertial fusion}

Plasma equation of state (EOS), i.e., ionization state and the relationship between specific internal energies and 
pressure,
can be complicated depending on the regime of plasma temperature and density as well as the plasma species or 
mixture of species.  In magnetic fusion plasmas, which typically are at low density and consist
of fully ionized hydrogen isotopes, EOS
is relatively straightforward, e.g., governed by
the ideal gas law $p=n_e kT_e + n_i k T_i$ with $n_e=n_i$.  On the other hand,
in inertial fusion and HED plasmas, which can have mixed-species plasmas with less than fully stripped
ions, EOS is a leading order issue complicated by atomic
physics, and significant effort has been made
and sophisticated capabilities developed to model EOS\@.

The PLX railguns can generate low-to-moderate density plasmas using any species or mixture available in a 
compressed-gas bottle.  The density and temperature range (see Table~\ref{tab:parameters}) are such
that the mean charge state $\bar{Z}$ is rapidly varying, and local-thermodynamic-equilibrium (LTE) and non-LTE models
predict notably different EOS\@.  PLX plasmas, whether single- or merging-jet, offer a platform for
measuring and constraining EOS in a plasma regime that has not received much detailed attention.  

\subsection{Interaction of plasma flow with vacuum magnetic field or magnetized plasma}

Studies of plasma flow across a magnetic field \citep[e.g.,][]{baker65},
or into another magnetized plasma \citep[e.g.,][]{liu11nf} have a long history, and
have been motivated by both basic plasma physics and applications such as tokamak fueling \citep[e.g.,][]{perkins88}.
With our in-chamber HH coil, we can study supersonic plasma jet interactions with a vacuum magnetic field.
In addition, because our plasma jets each consist of multiple ``blobs" due to the ringing railgun current, we can study
the interaction of trailing blobs with stagnated magnetized plasma formed from the merged leading blobs.

\section{Summary of major results}
\label{sec:results}

\subsection{Characterization of single-jet parameters and evolution}
\label{sec:single-jet}

The first set of experiments that we performed on PLX focused on the detailed
characterization of the plasma parameters, profiles, and evolution over $\approx 1$~m of propagation of a 
single argon plasma jet launched by a plasma railgun \citep{hsu12b}.  These
experiments were done to help us understand
and interpret planned, subsequent experiments based on the merging of two or more plasma jets, and also to provide
accurate initial conditions for numerical modeling of experiments using these plasma jets.
These experiments also contributed uniquely to the plasma railgun literature, most of which has focused on in-bore
characterization of the plasma armature as a pusher for launching solid projectiles to high velocities
\citep[see][for example]{batteh91}.
Table~\ref{tab:parameters} summarizes the measured (or experimentally inferred) argon 
plasma jet parameters
at the exit of the railgun nozzle ($Z\approx 2$~cm) and downstream ($Z\approx 41$~cm), as well as
the expected range of values for some of the parameters, although the latter were not systematically measured in the
single-jet studies.

The plasma jet is essentially unmagnetized after a few tens of centimeters of propagation (at 
a typical speed of 3--7~cm/$\mu$s) due to fast resistive decay of
the magnetic field that is embedded in the jet as it exits the railgun bore.
Using magnetic pickup coils mounted on the exterior surface of the railgun nozzle, we have
measured \citep{merritt14} the decay time of the transverse (relative to the jet propagation) component of the
magnetic field (of order 750~G near the railgun nozzle exit for
peak railgun current of order 200~kA) to be $\approx 5.6$~$\mu$s, which is consistent with the expected magnetic diffusion time
estimated using the plasma jet parameters and radial scale lengths.
Thus, for the jet-merging studies described in Secs.~\ref{sec:oblique} and \ref{sec:head-on},
the jet kinetic energy density ($\rho v^2/2$) is much greater (by a factor of 50--$10^4$) than the magnetic
energy density ($B^2/2\mu_0$) at the time of initial jet merging.

\begin{table}
\begin{center}
\def~{\hphantom{0}}
\begin{tabular}{lccc}
parameter & $Z\approx 2$~cm & $Z\approx 41$~cm & range\\[3 pt]
ion+neutral density $n_{\rm tot}$~(cm$^{-3}$) & $2\times 10^{16}$ & $2\times 10^{15}$ & $10^{14}$--$10^{17}$\\
electron temperature $T_e$ (eV) & 1.4 & 1.4 & 1--3\\
jet velocity $V$ (km/s) & 30 & 30 & 30--100\\
mean charge $\bar{Z}$ & 0.96 & 0.94 & $\sim 1$\\
jet length $L$ (cm) & 20 & 45\\
jet diameter $D$ (cm) & 5 & 10--20\\
\end{tabular}
\caption{Measured and achievable range of railgun-driven argon plasma jet parameters.  Hydrogen and helium
jets have similar values.}
\label{tab:parameters}
\end{center}
\end{table}

\subsection{Evidence for collisional shock formation between obliquely merging plasma jets}
\label{sec:oblique}

Following the single-jet experiments, we conducted experiments on the oblique merging of two plasma jets 
\citep{merritt13,merritt14}, in order to gain better understanding of
the essential underlying physical process for plasma liner formation via merging
plasma jets.  This work led to the identification and characterization of the stagnation layer between
obliquely merging jets and the demonstration that the stagnation layer morphology is consistent
with that of a collisional oblique plasma shock \citep{merritt13,merritt14}.  

We obtained temporally and spatially resolved survey-spectrometer and interferometer measurements
of the stagnation layer, along with sequences of time-gated visible CCD images.  Based
on experimentally inferred plasma parameters, the inter-jet ion mean free path were less than
the stagnation layer thickness, i.e., the oblique jet-merging was in a collisional regime.  Detailed analyses of the
data from all three diagnostics allowed us to arrive at the following key results
\citep{merritt13,merritt14}:  demonstration
that (a)~the observed stagnation layer morphology is consistent with that expected of a collisional oblique
plasma shock, (b)~the peak density exceeds that of simple plasma interpenetration but is about a factor
of 2 smaller than the peak density predicted by hydrodynamic shock theory, and (c)~the observed gradient scale
length at the postulated shock boundary is consistent with two-fluid plasma theory predictions of shock layer
thickness.

\subsection{Transition from collisionless interpenetration to collisional stagnation during head-on jet merging}
\label{sec:head-on}

In working toward a more collisionless regime for the counter-streaming ions between the merging
jets, we performed a set of head-on jet-merging experiments
in which the interaction between jets
began with collisionless interpenetration but then transitioned to collisional stagnation
due to a rising $\bar{Z}$ \citep{moser14}.
%The latter was attributed to electron frictional heating by the ions of one jet
%slowing down on the electrons of the opposing jet.
This observation demonstrates a clear way by which
an initially collisionless interaction between colliding plasmas can evolve toward a collisional one, and provides
data that can be used to validate physics models of plasma collisionality in the presence of complex EOS\@.
The observation of unmagnetized,
supersonic interpenetration also establishes a regime in which neither macro- nor micro-instabilities are
present or strong enough to introduce collisionless shock effects.

\subsection{Observation of Rayleigh-Taylor-instability growth and progression to longer wavelength}

As  mentioned earlier, our ringing railgun current produces several serial plasma ``blobs," each
separated in time by 20--30~$\mu$s.  The leading blob of each jet each compresses the applied magnetic field and
stagnates in the center of the chamber upon merging with the opposing jet.
We observed one trailing blob decelerating against the stagnated magnetized
plasma in the center (formed by the merged leading blobs).  Images from a multiple-frame CCD camera
(Invisible Vision model UHSi 12/24) revealed
the growth of Rayleigh-Taylor-like fingers at the leading edge of this trailing blob as well as the evolution toward
longer wavelength of these fingers.  Preliminary instability-growth-rate estimates 
and magnetohydrodynamic (MHD) simulations suggest
that our observations are consistent with Rayleigh-Taylor instability of this decelerating interface.  The progression
toward longer mode wavelength could be due to a combination of magnetic stabilization and viscous damping.
Detailed analysis is ongoing, and these results will be reported elsewhere.

\subsection{Ongoing studies of magnetized head-on jet merging to study cosmically relevant collisionless
shocks}
\label{sec:magnetized}

Ongoing experiments on PLX are aimed at forming and studying ``cosmically relevant" collisionless shocks, as
described by \citet{drake00}, who argues that such experiments must meet a number of competing constraints
on magnetization, plasma $\beta$, Alfv\'en Mach number $M_A$, and collisionality.  
Table~\ref{tab:criteria} summarizes the original design parameters for collisionless shock studies on 
PLX using head-on-merging plasma jets.  Hybrid particle-in-cell (PIC) and fully kinetic PIC simulations
of head-on merging jets in this parameter regime confirmed that we would be in a collisionless regime
and that magnetized shocks should form \citep{thoma13}.  However, the level of impurities in our railgun-driven
jets was higher than expected \citep{hsu12b,merritt14,moser14}, leading to an unanticipated rise in $\bar{Z}$ (even in hydrogen
experiments), causing a reduction in the counter-streaming ion-ion mean free path that scales
as $\bar{Z}^{-4}$, and a transition to a collisional regime
(described in Sec.~\ref{sec:head-on}) at the originally chosen jet velocities.  Thus, ongoing work is aimed
at reaching collisionless regimes for
the counter-streaming ions by achieving even higher counter-streaming jet velocities to overcome the rising-$\bar{Z}$ 
effect.  Failing this, we may need to upgrade to coaxial guns \citep{witherspoon09}, which are
expected to have far lower levels of impurities.

\begin{table}
\begin{center}
\def~{\hphantom{0}}
\begin{tabular}{lcc}
parameter & jets at collision & post-shock\\
species & H$^+$ & H$^+$ \\
density (cm$^{-3}$) & $3\times 10^{14}$ & $1.2\times 10^{15}$\\
temperature (eV) & 1 & $T_i=139$\\
speed (km/s) & $V_{\rm jet}=100$ & $V_{\rm shock}=167$\\
magnetic field (G) & 300 (applied) &1200 \\
length $L$ (cm) & 50 & \\
radius $R$ (cm) & 15 & \\
%background vacuum pressure (Torr) & $10^{-6}$ & $10^{-6}$ \\
%shock speed $V_s$ (km/s) & & 167\\
%post-shock $T_i$ (eV) & 139 & N/A \\
\hline
criterion & estimate for proposed experiment & \\
 & (post-shock values) & \\
$2R/\rho_i\gg 1$ & 30 & \\
$\beta > 1$ & 4.7 & \\
$M_A > 1$ & 2.2 & \\
$\lambda_{i}/\rho_i \gg 1$ & 28 & \\
$R_M \gg 1$ & $\sim 100$ (for $T_e\sim 10$~eV) \\
%$\lambda_{in} > L$ & $>2$ (for 1\% neutrals in jet) & \\
$\omega_{ci}\tau_{exp} \gg 1$ & 34 & \\
$L/(c/\omega_{pi})\gg 1$ & 76 & \\
\end{tabular}
\caption{Proposed reference experimental values, and evaluation of
relevant physics criteria against
those of \citet{drake00} for a cosmically relevant collisionless shock
experiment on PLX\@.  Here, $\rho_i$ is the thermal ion gyro-radius, $\beta$ the ratio
of plasma thermal-to-magnetic pressure, $M_A$ the Alfv\'en Mach number, $\lambda_i$
the ion mean free path, $R_M$ the magnetic Reynold's number, 
$\omega_{ci}$ the ion-cyclotron frequency, $\tau_{exp}$ the shock-transit time, and $c/\omega_{pi}$
the ion inertial length.  All values given are for a quasi-perpendicular shock, and all speeds are
in the laboratory frame.}
\label{tab:criteria}
\end{center}
\end{table}

\section{Future opportunities}
\label{sec:opportunities}

Future opportunities on PLX with two plasma guns include further and more detailed studies of
all the topics discussed in Secs.~\ref{sec:physics} and \ref{sec:results}.  Upgraded coaxial
guns \citep[e.g.,][]{witherspoon09} would reduce impurities and allow for higher velocities than railguns.
In addition, the plasma shock and interpenetration data offer a rare opportunity to validate collisionality,
EOS, and shock-handling physics models employed in MHD, multi-fluid, and kinetic plasma codes.

If PLX is upgraded to have many more guns, e.g., 36 or 60, then the merging and
spherical convergence of all the jets
could produce cm-, $\mu$s-, and Mbar-scale plasmas, enabling unique
fundamental physics studies of many HED topics (as described in the U.S. Dept.\ of Energy Report of the Workshop
on High
Energy Density Laboratory Physics Research Needs, Nov.~15--18, 2009) and
the exploration of a standoff MIF driver for fusion energy \citep{hsu12a}.

\section*{Acknowledgments}
We thank Drs.~Y. C. F. Thio and J. T. Cassibry for many useful conversations and Drs.~Glen Wurden
and Thomas Intrator for loaning numerous items of laboratory and diagnostic equipment,
especially the multiple-frame CCD camera.
This work was supported by the Office of Science (Office of Fusion Energy Sciences)
and the LANL Laboratory Directed Research and Development (LDRD) Program
under U.S. Department of Energy contract no.~DE-AC52-06NA25396.

\bibliographystyle{jpp}

%\bibliography{ms}

\begin{thebibliography}{47}
\expandafter\ifx\csname natexlab\endcsname\relax\def\natexlab#1{#1}\fi

\bibitem[Adams {\em et~al.\/}(2012)Adams, Lynn, Gilmore, Merritt, Moser \&
  Hsu]{adams12}
{\sc Adams, C.~S., Lynn, A.~G., Gilmore, M.~A., Merritt, E.~C., Moser, A.~L. \&
  Hsu, S.~C.} 2012 Schlieren imaging diagnostic for a collisionless shock
  experiment. {\em Bull.\ Amer.\ Phys.\ Soc.\/} {\bf 57}, 130.

\bibitem[Awe {\em et~al.\/}(2011)Awe, Adams, Davis, Hanna, Hsu \&
  Cassibry]{awe11}
{\sc Awe, T.~J., Adams, C.~S., Davis, J.~S., Hanna, D.~S., Hsu, S.~C. \&
  Cassibry, J.~T.} 2011 One-dimensional radiation-hydrodynamic scaling studies
  of imploding spherical plasma liners. {\em Phys.\ Plasmas\/} {\bf 18},
  072705.

\bibitem[Baker \& Hammel(1965)]{baker65}
{\sc Baker, D.~A. \& Hammel, J.~E.} 1965 Experimental studies of the
  penetration of a plasma stream into a transverse magnetic field. {\em Phys.\
  Fluids\/} {\bf 8}, 713.

\bibitem[Batteh(1991)]{batteh91}
{\sc Batteh, J.~H.} 1991 Review of armature research. {\em IEEE Trans.\
  Magn.\/} {\bf 27}, 224.

\bibitem[Bellan {\em et~al.\/}(2005)Bellan, You \& Hsu]{bellan05}
{\sc Bellan, P.~M., You, S. \& Hsu, S.~C.} 2005 Simulating astrophysical jets
  in laboratory experiments. {\em Astrophys.\ Space Sci.\/} {\bf 298}, 203.

\bibitem[Casanova {\em et~al.\/}(1991)Casanova, Larroche \& Matte]{casanova91}
{\sc Casanova, M., Larroche, O. \& Matte, J.-P.} 1991 Kinetic simulation of a
  collisional shock wave in a plasma. {\em Phys.\ Rev.\ Lett.\/} {\bf 67},
  2143.

\bibitem[Cassibry {\em et~al.\/}(2013)Cassibry, Stanic \& Hsu]{cassibry13}
{\sc Cassibry, J.~T., Stanic, M. \& Hsu, S.~C.} 2013 Ideal hydrodynamic scaling
  relations for a stagnated imploding spherical plasma liner formed by an array
  of merging plasma jets. {\em Phys.\ Plasmas\/} {\bf 20}, 032706.

\bibitem[Davis {\em et~al.\/}(2012)Davis, Hsu, Golovkin, MacFarlane \&
  Cassibry]{davis12}
{\sc Davis, J.~S., Hsu, S.~C., Golovkin, I.~E., MacFarlane, J.~J. \& Cassibry,
  J.~T.} 2012 One-dimensional radiation-hydrodynamic simulations of imploding
  spherical plasma liners with detailed equation-of-state modeling. {\em Phys.\
  Plasmas\/} {\bf 19}, 102701.

\bibitem[Degnan {\em et~al.\/}(2013)Degnan, Amdahl, Domonkos, Lehr, Grabowski,
  Robinson, Ruden, White, Wurden, Intrator, Sears, Weber, Waganaar, Frese,
  Frese, Camacho, Coffey, Makhin, Roderick, Gale, Kostora, Lerma, McCullough,
  Sommars, Kiuttu, Bauer, Fuelling, Siemon, Lynn \& Turchi]{degnan13}
{\sc Degnan, J.H., Amdahl, D.J., Domonkos, M., Lehr, F.M., Grabowski, C.,
  Robinson, P.R., Ruden, E.L., White, W.M., Wurden, G.A., Intrator, T.P.,
  Sears, J., Weber, T., Waganaar, W.J., Frese, M.H., Frese, S.D., Camacho,
  J.F., Coffey, S.K., Makhin, V., Roderick, N.F., Gale, D.G., Kostora, M.,
  Lerma, A., McCullough, J.L., Sommars, W., Kiuttu, G.F., Bauer, B., Fuelling,
  S.R., Siemon, R.E., Lynn, A.G. \& Turchi, P.J.} 2013 Recent magneto-inertial
  fusion experiments on the field reversed configuration heating experiment.
  {\em Nucl.\ Fusion\/} {\bf 53}, 093003.

\bibitem[Drake(2000)]{drake00}
{\sc Drake, R.~P.} 2000 The design of laboratory experiments to produce
  collisionless shocks of cosmic relevance. {\em Phys.\ Plasmas\/} {\bf 7},
  4690.

\bibitem[Drake(2006)]{drake06}
{\sc Drake, R.~P.} 2006 {\em High-Energy-Density-Physics\/}. Berlin: Springer.

\bibitem[Fiksel {\em et~al.\/}(2014)Fiksel, Fox, Bhattacharjee, Barnak, Chang,
  Germaschewski, Hu \& Nilson]{fiksel14}
{\sc Fiksel, G., Fox, W., Bhattacharjee, A., Barnak, D.~H., Chang, P.-Y.,
  Germaschewski, K., Hu, S.~X. \& Nilson, P.~M.} 2014 Magnetic reconnection
  between colliding magnetized laser-produced plasma plumes. {\em Phys.\ Rev.\
  Lett.\/} {\bf 113}, 105003.

\bibitem[Fox {\em et~al.\/}(2013)Fox, Fiksel, Bhattacharjee, Chang,
  Germaschewski, Hu \& Nilson]{fox13}
{\sc Fox, W., Fiksel, G., Bhattacharjee, A., Chang, P.-Y., Germaschewski, K.,
  Hu, S.~X. \& Nilson, P.~M.} 2013 Filamentation instability of
  counterstreaming laser-driven plasmas. {\em Phys.\ Rev.\ Lett.\/} {\bf 111},
  225002.

\bibitem[Gourdain \& Seyler(2013)]{gourdain13}
{\sc Gourdain, P.-A. \& Seyler, C.~E.} 2013 Impact of the hall effect on
  high-energy-density plasma jets. {\em Phys. Rev. Lett.\/} {\bf 110}, 015002.

\bibitem[Haas {\em et~al.\/}(2011)Haas, Bott, Kim, Mariscal, Madden, Eshaq,
  Ueda, \mbox{Collins~IV}, Gunasekera, Beg, Chittenden, Niasse \&
  Jennings]{haas11}
{\sc Haas, D.~M., Bott, S.~C., Kim, J., Mariscal, D.~A., Madden, R.~E., Eshaq,
  Y., Ueda, U., \mbox{Collins~IV}, G., Gunasekera, K., Beg, F.~N., Chittenden,
  J.~P., Niasse, N. \& Jennings, C.~A.} 2011 Supersonic jet formation and
  propagation in x-pinches. {\em Astrophys.\ Space Sci.\/} {\bf 336}, 33.

\bibitem[Hsu(2009)]{hsu09}
{\sc Hsu, S.~C.} 2009 Technical summary of the first \mbox{U.S.} plasma jet
  workshop. {\em J. Fusion Energy\/} {\bf 28}, 246.

\bibitem[Hsu {\em et~al.\/}(2012{\natexlab{{\em a\/}}})Hsu, Awe, Brockington,
  Case, Cassibry, Kagan, Messer, Stanic, Tang, Welch \& Witherspoon]{hsu12a}
{\sc Hsu, S.~C., Awe, T.~J., Brockington, S., Case, A., Cassibry, J.~T., Kagan,
  G., Messer, S.~J., Stanic, M., Tang, X., Welch, D.~R. \& Witherspoon, F.~D.}
  2012{\natexlab{{\em a\/}}} Spherically imploding plasma liners as a standoff
  driver for magnetoinertial fusion. {\em IEEE Trans.\ Plasma Sci.\/} {\bf 40},
  1287.

\bibitem[Hsu \& Bellan(2005)]{hsu05}
{\sc Hsu, S.~C. \& Bellan, P.~M.} 2005 On the jets, kinks, and spheromaks
  formed by a planar magnetized coaxial gun. {\em Phys. Plasmas\/} {\bf 12},
  032103.

\bibitem[Hsu {\em et~al.\/}(2012{\natexlab{{\em b\/}}})Hsu, Merritt, Moser,
  Awe, Brockington, Davis, Adams, Case, Cassibry, Dunn, Gilmore, Lynn, Messer
  \& Witherspoon]{hsu12b}
{\sc Hsu, S.~C., Merritt, E.~C., Moser, A.~L., Awe, T.~J., Brockington, S.
  J.~E., Davis, J.~S., Adams, C.~S., Case, A., Cassibry, J.~T., Dunn, J.~P.,
  Gilmore, M.~A., Lynn, A.~G., Messer, S.~J. \& Witherspoon, F.~D.}
  2012{\natexlab{{\em b\/}}} Experimental characterization of railgun-driven
  supersonic plasma jets motivated by high energy density physics applications.
  {\em Phys.\ Plasmas\/} {\bf 19}, 123514.

\bibitem[Intrator {\em et~al.\/}(2004)Intrator, Zhang, Degnan, Furno,
  Grabowski, Hsu, Ruden, Sanchez, Taccetti, Tuszewski, Waganaar \&
  Wurden]{intrator04}
{\sc Intrator, T., Zhang, S.~Y., Degnan, J.~H., Furno, I., Grabowski, C., Hsu,
  S.~C., Ruden, E.~L., Sanchez, P.~G., Taccetti, J.~M., Tuszewski, M.,
  Waganaar, W.~J. \& Wurden, G.~A.} 2004 A high density field reversed
  configuration \mbox{(FRC)} target for magnetized target fusion: First
  internal profile measurements of a high density \mbox{FRC}. {\em Phys.\
  Plasmas\/} {\bf 11}, 2580--2585.

\bibitem[Jaffrin \& Probstein(1964)]{jaffrin64}
{\sc Jaffrin, M.~Y. \& Probstein, R.~F.} 1964 Structure of a plasma shock wave.
  {\em Phys.\ Fluids\/} {\bf 7}, 1658.

\bibitem[Ji {\em et~al.\/}(1991)Ji, Toyama, Yamagishi, Shinohara, Fujisawa \&
  Miyamoto]{ji91}
{\sc Ji, H., Toyama, H., Yamagishi, K., Shinohara, S., Fujisawa, A. \&
  Miyamoto, K.} 1991 Probe measurements in the \mbox{REPUTE-1} reversed field
  pinch. {\em Rev.\ Sci.\ Instrum.\/} {\bf 62}, 2326.

\bibitem[Kirkpatrick {\em et~al.\/}(1995)Kirkpatrick, Lindemuth \&
  Ward]{kirkpatrick95}
{\sc Kirkpatrick, R.~C., Lindemuth, I.~R. \& Ward, M.~S.} 1995 Magnetized
  target fusion: An overview. {\em Fusion Tech.\/} {\bf 27}, 201.

\bibitem[Knapp \& Kirkpatrick(2014)]{knapp14}
{\sc Knapp, C.~E. \& Kirkpatrick, R.~C.} 2014 Possible energy gain for a
  plasma-liner-driven magneto-inertial fusion concept. {\em Phys.\ Plasmas\/}
  {\bf 21}, 070701.

\bibitem[Li {\em et~al.\/}(2013)Li, Ryutov, Hu, Rosenberg, Zylstra, S\'eguin,
  Frenje, Casey, Gatu~Johnson, Manuel, Rinderknecht, Petrasso, Amendt, Park,
  Remington, Wilks, Betti, Froula, Knauer, Meyerhofer, Drake, Kuranz, Young \&
  Koenig]{li13}
{\sc Li, C.~K., Ryutov, D.~D., Hu, S.~X., Rosenberg, M.~J., Zylstra, A.~B.,
  S\'eguin, F.~H., Frenje, J.~A., Casey, D.~T., Gatu~Johnson, M., Manuel, M.
  J.-E., Rinderknecht, H.~G., Petrasso, R.~D., Amendt, P.~A., Park, H.~S.,
  Remington, B.~A., Wilks, S.~C., Betti, R., Froula, D.~H., Knauer, J.~P.,
  Meyerhofer, D.~D., Drake, R.~P., Kuranz, C.~C., Young, R. \& Koenig, M.} 2013
  Structure and dynamics of colliding plasma jets. {\em Phys. Rev. Lett.\/}
  {\bf 111}, 235003.

\bibitem[Lindemuth \& Kirkpatrick(1983)]{lindemuth83}
{\sc Lindemuth, I.~R. \& Kirkpatrick, R.~C.} 1983 Parameter space for
  magnetized fuel targets in inertial confinement fusion. {\em Nucl.\ Fusion\/}
  {\bf 23}, 263.

\bibitem[Lindemuth \& Siemon(2009)]{lindemuth09a}
{\sc Lindemuth, I.~R. \& Siemon, R.~E.} 2009 The fundamental parameter space of
  controlled thermonuclear fusion. {\em Amer.\ J.\ Phys.\/} {\bf 77}, 407.

\bibitem[Liu \& Hsu(2011)]{liu11nf}
{\sc Liu, W. \& Hsu, S.~C.} 2011 Ideal magnetohydrodynamic simulations of
  unmagnetized dense plasma jet injection into a hot strongly magnetized
  plasma. {\em Nucl.\ Fusion\/} {\bf 51}, 073026.

\bibitem[Lynn {\em et~al.\/}(2010)Lynn, Merritt, Gilmore, Hsu, Witherspoon \&
  Cassibry]{lynn10}
{\sc Lynn, A.~G., Merritt, E., Gilmore, M., Hsu, S.~C., Witherspoon, F.~D. \&
  Cassibry, J.~T.} 2010 Diagnostics for the \mbox{Plasma} \mbox{Liner}
  \mbox{Experiment}. {\em Rev.\ Sci.\ Instrum.\/} {\bf 81}, 10E115.

\bibitem[Merritt {\em et~al.\/}(2012{\natexlab{{\em a\/}}})Merritt, Lynn,
  Gilmore \& Hsu]{merritt12a}
{\sc Merritt, E.~C., Lynn, A.~G., Gilmore, M.~A. \& Hsu, S.~C.}
  2012{\natexlab{{\em a\/}}} Multi-chord fiber-coupled interferometer with a
  long coherence length laser. {\em Rev.\ Sci.\ Instrum.\/} {\bf 83}, 033506.

\bibitem[Merritt {\em et~al.\/}(2012{\natexlab{{\em b\/}}})Merritt, Lynn,
  Gilmore, Thoma, Loverich \& Hsu]{merritt12b}
{\sc Merritt, E.~C., Lynn, A.~G., Gilmore, M.~A., Thoma, C., Loverich, J. \&
  Hsu, S.~C.} 2012{\natexlab{{\em b\/}}} Multi-chord fiber-coupled
  interferometry of supersonic plasma jets. {\em Rev.\ Sci.\ Instrum.\/} {\bf
  83}, 10D523.

\bibitem[Merritt {\em et~al.\/}(2014)Merritt, Moser, Hsu, Adams, Dunn, Holgado
  \& Gilmore]{merritt14}
{\sc Merritt, E.~C., Moser, A.~L., Hsu, S.~C., Adams, C.~S., Dunn, J.~P.,
  Holgado, A.~M. \& Gilmore, M.} 2014 Experimental evidence for collisional
  shock formation via two obliquely merging supersonic plasma jets. {\em Phys.\
  Plasmas\/} {\bf 21}, 055703.

\bibitem[Merritt {\em et~al.\/}(2013)Merritt, Moser, Hsu, Loverich \&
  Gilmore]{merritt13}
{\sc Merritt, E.~C., Moser, A.~L., Hsu, S.~C., Loverich, J. \& Gilmore, M.}
  2013 Experimental characterization of the stagnation layer between two
  obliquely merging supersonic plasma jets. {\em Phys.\ Rev.\ Lett.\/} {\bf
  111}, 085003.

\bibitem[Moser \& Bellan(2012)]{moser12}
{\sc Moser, A.~L. \& Bellan, P.~M.} 2012 Magnetic reconnection from a
  multiscale instability cascade. {\em Nature\/} {\bf 482}, 379.

\bibitem[Moser \& Hsu(2014)]{moser14}
{\sc Moser, A.~L. \& Hsu, S.~C.} 2014 Observation of ionization-mediated
  transition from collisionless interpenetration to collisional stagnation
  during merging of two supersonic plasmas. {\em submitted{\rm ;}\/}
  \mbox{http://arxiv.org/abs/1405.2286}.

\bibitem[Perkins {\em et~al.\/}(1988)Perkins, Ho \& Hammer]{perkins88}
{\sc Perkins, L.~J., Ho, S.~K. \& Hammer, J.~H.} 1988 Deep penetration fuelling
  of reactor-grade tokamak plasmas with accelerated compact toroids. {\em
  Nucl.\ Fus.\/} {\bf 28}, 1365.

\bibitem[Romero-Talam\'as {\em et~al.\/}(2004)Romero-Talam\'as, Bellan \&
  Hsu]{romero04}
{\sc Romero-Talam\'as, C.~A., Bellan, P.~M. \& Hsu, S.~C.} 2004 Multielement
  magnetic probe using commercial chip inductors. {\em Rev.\ Sci.\ Instrum.\/}
  {\bf 75}, 2664.

\bibitem[Ross {\em et~al.\/}(2013)Ross, Park, Berger, Divol, Kugland, Rozmus,
  Ryutov \& Glenzer]{ross13}
{\sc Ross, J.~S., Park, H.-S., Berger, R., Divol, L., Kugland, N.~L., Rozmus,
  W., Ryutov, D. \& Glenzer, S.~H.} 2013 Collisionless coupling of ion and
  electron temperatures in counterstreaming plasma flows. {\em Phys.\ Rev.\
  Lett.\/} {\bf 110}, 145005.

\bibitem[Sagdeev \& Kennel(1991)]{sagdeev91}
{\sc Sagdeev, R.~Z. \& Kennel, C.~F.} 1991 Collisionless shock waves. {\em
  Sci.\ Amer.\/} {\bf 264}, 106.

\bibitem[Santarius(2012)]{santarius12}
{\sc Santarius, J.~F.} 2012 Compression of a spherically symmetric
  deuterium-tritium plasma liner onto a magnetized deuterium-tritium target.
  {\em Phys.\ Plasmas\/} {\bf 19}, 072705.

\bibitem[Settles(2001)]{settles01}
{\sc Settles, G.~S.} 2001 {\em Schlieren and Shadowgraph Techniques\/}. New
  York: Springer.

\bibitem[Slutz {\em et~al.\/}(2010)Slutz, Herrmann, Vesey, Sefkow, Sinars,
  Rovang, Peterson \& Cuneo]{slutz10}
{\sc Slutz, S.~A., Herrmann, M.~C., Vesey, R.~A., Sefkow, A.~B., Sinars, D.~B.,
  Rovang, D.~C., Peterson, K.~J. \& Cuneo, M.~E.} 2010 Pulsed-power-driven
  cylindrical liner implosions of laser preheated fuel magnetized with an axial
  field. {\em Phys.\ Plasmas\/} {\bf 17}, 056303.

\bibitem[Swadling {\em et~al.\/}(2014)Swadling, Lebedev, Harvey-Thompson,
  Rozmus, Burdiak, Suttle, Patankar, Smith, Bennett, Hall, Suzuki-Vidal \&
  Yuan]{swadling14}
{\sc Swadling, G.~F., Lebedev, S.~V., Harvey-Thompson, A.~J., Rozmus, W.,
  Burdiak, G.~C., Suttle, L., Patankar, S., Smith, R.~A., Bennett, M., Hall,
  G.~N., Suzuki-Vidal, F. \& Yuan, J.} 2014 Interpenetration, deflection, and
  stagnation of cylindrically convergent magnetized supersonic tungsten plasma
  flows. {\em Phys.\ Rev.\ Lett.\/} {\bf 113}, 035003.

\bibitem[Thio {\em et~al.\/}(2001)Thio, Knapp, Kirkpatrick, Siemon \&
  Turchi]{thio01}
{\sc Thio, Y. C.~F., Knapp, C.~E., Kirkpatrick, R.~C., Siemon, R.~E. \& Turchi,
  P.~J.} 2001 A physics exploratory experiment on plasma liner formation. {\em
  J. Fusion Energy\/} {\bf 20}, 1.

\bibitem[Thio {\em et~al.\/}(1999)Thio, Panarella, Kirkpatrick, Knapp, Wysocki,
  Parks \& Schmidt]{thio99}
{\sc Thio, Y. C.~F., Panarella, E., Kirkpatrick, R.~C., Knapp, C.~E., Wysocki,
  F., Parks, P. \& Schmidt, G.} 1999 Magnetized target fusion in a spheroidal
  geometry with standoff drivers. In {\em Current Trends in International
  Fusion Research--Proceedings of the Second International Symposium\/} (ed.
  E.~Panarella), p. 113. Ottawa: National Research Council of Canada.

\bibitem[Thoma {\em et~al.\/}(2013)Thoma, Welch \& Hsu]{thoma13}
{\sc Thoma, C., Welch, D.~R. \& Hsu, S.~C.} 2013 Particle-in-cell simulations
  of collisionless shock formation via head-on merging of two laboratory
  supersonic plasma jets. {\em Phys.\ Plasmas\/} {\bf 20}, 082128.

\bibitem[Witherspoon {\em et~al.\/}(2009)Witherspoon, Case, Messer,
  \mbox{Bomgardner~II}, Phillips, Brockington \& Elton]{witherspoon09}
{\sc Witherspoon, F.~D., Case, A., Messer, S.~J., \mbox{Bomgardner~II}, R.,
  Phillips, M.~W., Brockington, S. \& Elton, R.} 2009 A contoured gap coaxial
  plasma gun with injected plasma armature. {\em Rev.\ Sci.\ Instrum.\/} {\bf
  80}, 083506.

\end{thebibliography}

\end{document}